\documentclass[sigconf]{acmart}
\AtBeginDocument{%
  }

\acmConference{CHI 2025 Everyday AR through AI-in-the-Loop Workshop}
\begin{document}

\title{Augmenting Teamwork through AI Agents as Spatial Collaborators}

\author{Mariana Fern\'andez-Espinosa}
\email{mferna23@nd.edu}
\orcid{0009-0004-1116-2002}
\affiliation{%
  \institution{University of Notre Dame}
  \city{Notre Dame}
  \state{Indiana}
  \country{USA}
}

\author{Diego G\'omez-Zar\'a}
\email{dgomezara@nd.edu}
\orcid{0000-0002-4609-6293}
\affiliation{%
  \institution{University of Notre Dame}
  \city{Notre Dame}
  \state{Indiana}
  \country{USA}
}


\begin{abstract}
As Augmented Reality (AR) and Artificial Intelligence (AI) continue to converge, new opportunities emerge for AI agents to actively support human collaboration in immersive environments. While prior research has primarily focused on dyadic human-AI interactions, less attention has been given to Human-AI Teams (HATs) in AR, where AI acts as an adaptive teammate rather than a static tool. This position paper takes the perspective of team dynamics and work organization to propose that AI agents in AR should not only interact with individuals but also recognize and respond to team-level needs in real time.

We argue that spatially aware AI agents should dynamically generate the resources necessary for effective collaboration, such as virtual blackboards for brainstorming, mental map models for shared understanding, and memory recall of spatial configurations to enhance knowledge retention and task coordination. This approach moves beyond predefined AI assistance toward context-driven AI interventions that optimize team performance and decision-making.

\end{abstract}






\maketitle

\section{Introduction}

% First sentence/paragraph using AR... (Social AR?) New advancements in AR are allowing multiple people to interact together...
% One main challenge continues to foster relationships and collaboration in separated places...
% Moreover, using AI as a social agent in AR continues unexplored...
% Solutions as NPC... however, current limitations...

Recent advancements in Augmented Reality (AR) devices, such as the Microsoft HoloLens \cite{MicrosoftHoloLens}, which utilizes gaze-based selection and speech-driven AI assistance, and Meta’s Orion \cite{MetaOrion}, which enables ubiquitous and context-aware interactions, have significantly expanded possibilities for collaborative human interactions \cite{li2019holodoc, zhao2024exploring}. These technologies are transforming how teams communicate and collaborate by blending digital and physical environments, enabling more immersive and dynamic team interactions.

Research in Human-Computer Interaction (HCI) has extensively examined how online communication systems impact team dynamics and has proposed diverse solutions to enhance productivity and efficiency in virtual meetings \cite{harris2019joining, tang2023perspectives, o2011blended}. One such approach is the integration of AI agents, designed to improve coordination, facilitate communication, and support participation. For instance, ``MeetingCoach,'' an automated meeting summarization system \cite{Airgram2025, Rewatch2025}, detects various behavioral cues in meetings and consolidates them into a post-meeting dashboard to improve understanding of meeting dynamics \cite{samrose2021meetingcoach}, and ``Dittos,'' a personalized AI agent that replaces team members when they are unavailable \cite{leong2024dittos}, demonstrate AI’s potential to enhance teamwork in digital settings.

However, AI agents still face significant challenges, particularly in adapting to real-world, multi-user collaboration settings \cite{paolo2024call, puig2020watch}. Most AI systems operate in text-based or static environments \cite{bovo2024embardiment, cohn2024can}, limiting their ability to understand spatial relationships, object affordances, and team interactions in dynamic, real-time contexts \cite{yan2023inherent}. This restricts AI’s effectiveness in facilitating fluid and adaptive collaboration in immersive environments for team meetings.

\subsection{AI Agents in Immersive Environments}
To address these challenges, XR researchers advocate for the design of AI agents that are deeply integrated into real-world settings, capable of observing, interacting with, and continuously learning from their surroundings and human collaborators in a dynamic and adaptive manner \cite{Hirzle2023WhenXR}. A promising application of this approach is the development of LLM-based Non-Player Character (NPC) agents, which have been deployed in video games \cite{korkiakoski5148461empirical, Kopel2018ImplementingAI}, training simulations \cite{Mergen2023MedicalTraining}, and virtual collaboration spaces \cite{Wan2024BuildingLLM}. These agents have promised to enhance user engagement by responding to real-time inputs, adapting to human behavior, and facilitating interactive experiences.

Despite these advancements, AI agents in immersive environments still have significant limitations. Current AI-driven systems engage primarily with individuals rather than supporting team-level collaboration, missing the opportunity to leverage capabilities in blending digital and physical environments. Additionally, most AI agents in immersive environments are designed as static assistants, unable to adjust their role dynamically based on team needs. Given that team meetings often involve both remote and in-person participants, AI agents also lack an understanding of proxemics (personal space, team positioning, movement patterns), which directly impacts how they engage in immersive settings.

As team meetings remain a fundamental aspect of organizational work, it is crucial to design AI agents that go beyond one-on-one interactions and instead focus on team behaviors, coordination patterns, and collective decision-making processes. Advancing AI-driven collaboration in immersive environments requires systems that can manage attention across multiple users, facilitate teamwork without disrupting workflows, and dynamically respond to team needs.

\subsection{The Need for AR-Specific AI Agents}
While the emerging field of Social VR has provided valuable insights into team behavior in immersive environments \cite{sayadi2024feeling}, limited research has explored these dynamics in Augmented Reality \cite{wang2019exploring}. Existing studies in VR suggest that immersive environments can influence team cohesion, engagement, and trust, but AR presents distinct challenges and affordances that remain largely unexplored \cite{vanderLand2011ModelingMetaverse}.

Our prior work \cite{fernandez2025breaking} examined the differences between teams collaborating using Virtual Reality and those collaborating face-to-face. Specifically, we examined how VR affordances influence team members' closeness to each other when forming the team. Our findings underscored that newcomers meeting in VR felt more included in the team than newcomers meeting face-to-face. Our work suggests that team members' perceptions and collaboration attitudes are shaped by the medium they interact in, rather than solely by interpersonal factors. Thus, we need to expand research beyond VR and into AR, which presents different interaction dynamics, spatial constraints, and opportunities for AI-driven facilitation.

\section{Future Work and Workshop Contribution}
In our future work, we aim to analyze and leverage AR’s spatial computing capabilities in combination with team dynamics analysis to design AI agents that enhance team meetings. These AI agents will function as active collaborators, dynamically generating and adapting supplemental materials to improve communication, coordination, and shared understanding. For instance, AI agents could create virtual blackboards for brainstorming, real-time subtitles for accessibility, or on-demand 3D models to help teams visualize complex concepts. By adapting to team needs in real-time, these AI agents can bridge gaps in understanding, facilitate problem-solving, and enhance collaborative decision-making in immersive AR environments.

\subsection{Workshop Contribution}

During the workshop, we seek to bring a team-centered perspective, advocating for the real-time design of AI agents for team meetings. We are particularly interested in discussing how AI agents can integrate into AR-based team workflows and contribute to more effective and dynamic collaboration. Additionally, we plan to present early findings from our research on AI agents in mixed reality, contributing to ongoing discussions within the AR and AI research communities.

Building on prior research that has recognized the challenges of virtual and in-person meetings, we propose expanding the discussion to AR environments and addressing the following key research questions:

\begin{itemize}
    \item RQ1: What are the appropriate designs for AI agents in AR that enhance usability, interaction, and effective team collaboration?
    \item RQ2: What spatial and proxemic factors should be considered in AR to improve team coordination and engagement?
    \item RQ3: How should AI agents be embodied in AR to focus on team-level interactions rather than interpersonal, one-on-one engagements?
    \item RQ4: When and how should AI agents intervene in team workflows within AR? How can they detect when users need assistance without disrupting collaboration?

\end{itemize}

By shifting the role of AI agents from predefined tools to responsive, context-aware teammates, we can unlock new possibilities for collaborative intelligence in AR environments. Addressing these research questions will be critical for designing AI agents that seamlessly integrate into team workflows, optimize interaction in immersive spaces, and ultimately enhance team performance and decision-making.

Through this discussion, we aim to advance the understanding of AI in AR not just as an interface, but as an active enabler of human-AI team collaboration.

\bibliographystyle{ACM-Reference-Format}
\bibliography{sample-base}


\begin{thebibliography}{25}

%%% the Web address.

\ifx \showCODEN    \undefined \def \showCODEN     #1{\unskip}     \fi
\ifx \showISBNx    \undefined \def \showISBNx     #1{\unskip}     \fi
\ifx \showISBNxiii \undefined \def \showISBNxiii  #1{\unskip}     \fi
\ifx \showISSN     \undefined \def \showISSN      #1{\unskip}     \fi
\ifx \showLCCN     \undefined \def \showLCCN      #1{\unskip}     \fi
\ifx \shownote     \undefined \def \shownote      #1{#1}          \fi
\ifx \showarticletitle \undefined \def \showarticletitle #1{#1}   \fi
\ifx \showURL      \undefined \def \showURL       {\relax}        \fi
% invisible to TeX
\providecommand\bibfield[2]{#2}
\providecommand\bibinfo[2]{#2}
\providecommand\natexlab[1]{#1}
\providecommand\showeprint[2][]{arXiv:#2}

\bibitem[{Airgram.io}(2025)]%
        {Airgram2025}
\bibfield{author}{\bibinfo{person}{{Airgram.io}}.} \bibinfo{year}{2025}\natexlab{}.
\newblock \bibinfo{booktitle}{\emph{Airgram | AI Assistant for Automated Meeting Notes \& Summary}}.
\newblock
\urldef\tempurl%
\url{https://www.airgram.io/}
\showURL{%
\tempurl}


\bibitem[Bovo et~al\mbox{.}(2024)]%
        {bovo2024embardiment}
\bibfield{author}{\bibinfo{person}{Riccardo Bovo}, \bibinfo{person}{Steven Abreu}, \bibinfo{person}{Karan Ahuja}, \bibinfo{person}{Eric~J Gonzalez}, \bibinfo{person}{Li-Te Cheng}, {and} \bibinfo{person}{Mar Gonzalez-Franco}.} \bibinfo{year}{2024}\natexlab{}.
\newblock \showarticletitle{Embardiment: an embodied ai agent for productivity in xr}.
\newblock \bibinfo{journal}{\emph{arXiv preprint arXiv:2408.08158}} (\bibinfo{year}{2024}).
\newblock


\bibitem[Cohn and Blackwell(2024)]%
        {cohn2024can}
\bibfield{author}{\bibinfo{person}{Anthony~G Cohn} {and} \bibinfo{person}{Robert~E Blackwell}.} \bibinfo{year}{2024}\natexlab{}.
\newblock \showarticletitle{Can Large Language Models Reason about the Region Connection Calculus?}
\newblock \bibinfo{journal}{\emph{arXiv preprint arXiv:2411.19589}} (\bibinfo{year}{2024}).
\newblock


\bibitem[Fernandez-Espinosa et~al\mbox{.}(2025)]%
        {fernandez2025breaking}
\bibfield{author}{\bibinfo{person}{Mariana Fernandez-Espinosa}, \bibinfo{person}{Kara Clouse}, \bibinfo{person}{Dylan Sellars}, \bibinfo{person}{Danny Tong}, \bibinfo{person}{Michael Bsales}, \bibinfo{person}{Sophonie Alcindor}, \bibinfo{person}{Timothy~D Hubbard}, \bibinfo{person}{Michael Villano}, {and} \bibinfo{person}{Diego G{\'o}mez-Zar{\'a}}.} \bibinfo{year}{2025}\natexlab{}.
\newblock \showarticletitle{Breaking the Familiarity Bias: Employing Virtual Reality Environments to Enhance Team Formation and Inclusion}.
\newblock \bibinfo{journal}{\emph{arXiv preprint arXiv:2502.09912}} (\bibinfo{year}{2025}).
\newblock


\bibitem[Harris et~al\mbox{.}(2019)]%
        {harris2019joining}
\bibfield{author}{\bibinfo{person}{Alexa~M Harris}, \bibinfo{person}{Diego G{\'o}mez-Zar{\'a}}, \bibinfo{person}{Leslie~A DeChurch}, {and} \bibinfo{person}{Noshir~S Contractor}.} \bibinfo{year}{2019}\natexlab{}.
\newblock \showarticletitle{Joining together online: the trajectory of CSCW scholarship on group formation}.
\newblock \bibinfo{journal}{\emph{Proceedings of the ACM on Human-Computer Interaction}} \bibinfo{volume}{3}, \bibinfo{number}{CSCW} (\bibinfo{year}{2019}), \bibinfo{pages}{1--27}.
\newblock


\bibitem[Hirzle et~al\mbox{.}(2023)]%
        {Hirzle2023WhenXR}
\bibfield{author}{\bibinfo{person}{Teresa Hirzle}, \bibinfo{person}{Florian M{\"u}ller}, \bibinfo{person}{Fiona Draxler}, \bibinfo{person}{Martin Schmitz}, \bibinfo{person}{Pascal Knierim}, {and} \bibinfo{person}{Kasper Hornb{\ae}k}.} \bibinfo{year}{2023}\natexlab{}.
\newblock \showarticletitle{When XR and AI Meet - A Scoping Review on Extended Reality and Artificial Intelligence}. In \bibinfo{booktitle}{\emph{Proceedings of the 2023 CHI Conference on Human Factors in Computing Systems}}. \bibinfo{publisher}{ACM}, \bibinfo{pages}{1--45}.
\newblock
\href{https://doi.org/10.1145/3544548.3581072}{doi:\nolinkurl{10.1145/3544548.3581072}}


\bibitem[Kopel and Smrz(2018)]%
        {Kopel2018ImplementingAI}
\bibfield{author}{\bibinfo{person}{Marek Kopel} {and} \bibinfo{person}{Pavel Smrz}.} \bibinfo{year}{2018}\natexlab{}.
\newblock \showarticletitle{Implementing AI for Non-player Characters in 3D Video Games}.
\newblock In \bibinfo{booktitle}{\emph{Advances in Intelligent Systems and Computing}}. Vol.~\bibinfo{volume}{659}. \bibinfo{publisher}{Springer}, \bibinfo{pages}{637--646}.
\newblock
\href{https://doi.org/10.1007/978-3-319-75417-8_57}{doi:\nolinkurl{10.1007/978-3-319-75417-8_57}}


\bibitem[Korkiakoski et~al\mbox{.}({[n.\,d.]})]%
        {korkiakoski5148461empirical}
\bibfield{author}{\bibinfo{person}{Mikko Korkiakoski}, \bibinfo{person}{Saeid Sheikhi}, \bibinfo{person}{Kalle Tapio}, \bibinfo{person}{Jesper Nyman}, \bibinfo{person}{Jussi Saariniemi}, {and} \bibinfo{person}{Panos Kostakos}.} \bibinfo{year}{[n.\,d.]}\natexlab{}.
\newblock \showarticletitle{An Empirical Evaluation of Ai-Powered Non-Player Characters’ Perceived Realism and Performance in Virtual Reality Environments}.
\newblock \bibinfo{journal}{\emph{Available at SSRN 5148461}} (\bibinfo{year}{[n.\,d.]}).
\newblock


\bibitem[Leong et~al\mbox{.}(2024)]%
        {leong2024dittos}
\bibfield{author}{\bibinfo{person}{Joanne Leong}, \bibinfo{person}{John Tang}, \bibinfo{person}{Edward Cutrell}, \bibinfo{person}{Sasa Junuzovic}, \bibinfo{person}{Gregory~Paul Baribault}, {and} \bibinfo{person}{Kori Inkpen}.} \bibinfo{year}{2024}\natexlab{}.
\newblock \showarticletitle{Dittos: Personalized, Embodied Agents That Participate in Meetings When You Are Unavailable}.
\newblock \bibinfo{journal}{\emph{Proceedings of the ACM on Human-Computer Interaction}} \bibinfo{volume}{8}, \bibinfo{number}{CSCW2} (\bibinfo{year}{2024}), \bibinfo{pages}{1--28}.
\newblock


\bibitem[Li et~al\mbox{.}(2019)]%
        {li2019holodoc}
\bibfield{author}{\bibinfo{person}{Zhen Li}, \bibinfo{person}{Michelle Annett}, \bibinfo{person}{Ken Hinckley}, \bibinfo{person}{Karan Singh}, {and} \bibinfo{person}{Daniel Wigdor}.} \bibinfo{year}{2019}\natexlab{}.
\newblock \showarticletitle{Holodoc: Enabling mixed reality workspaces that harness physical and digital content}. In \bibinfo{booktitle}{\emph{Proceedings of the 2019 CHI Conference on Human Factors in Computing Systems}}. \bibinfo{pages}{1--14}.
\newblock


\bibitem[Mergen et~al\mbox{.}(2023)]%
        {Mergen2023MedicalTraining}
\bibfield{author}{\bibinfo{person}{Marvin Mergen}, \bibinfo{person}{Anna Junga}, \bibinfo{person}{Benjamin Risse}, \bibinfo{person}{Dimitar Valkov}, \bibinfo{person}{Norbert Graf}, {and} \bibinfo{person}{Bernhard Marschall}.} \bibinfo{year}{2023}\natexlab{}.
\newblock \showarticletitle{Immersive training of clinical decision making with AI-driven virtual patients—a new VR platform called medical tr.AI.ning}.
\newblock \bibinfo{journal}{\emph{GMS Journal for Medical Education}} \bibinfo{volume}{40}, \bibinfo{number}{2} (\bibinfo{year}{2023}), \bibinfo{pages}{Doc19}.
\newblock
\href{https://doi.org/10.3205/zma001600}{doi:\nolinkurl{10.3205/zma001600}}


\bibitem[{Meta}(2024)]%
        {MetaOrion}
\bibfield{author}{\bibinfo{person}{{Meta}}.} \bibinfo{year}{2024}\natexlab{}.
\newblock \bibinfo{booktitle}{\emph{Introducing Orion: Our First True Augmented Reality Glasses}}.
\newblock
\urldef\tempurl%
\url{https://about.fb.com/news/2024/09/introducing-orion-our-first}
\showURL{%
\tempurl}


\bibitem[{Microsoft}(2024)]%
        {MicrosoftHoloLens}
\bibfield{author}{\bibinfo{person}{{Microsoft}}.} \bibinfo{year}{2024}\natexlab{}.
\newblock \bibinfo{booktitle}{\emph{Gaze and Commit - Designing for HoloLens}}.
\newblock
\urldef\tempurl%
\url{https://learn.microsoft.com/en-us/windows/mixed-reality/design/gaze-and-commit}
\showURL{%
\tempurl}
\newblock
\shownote{Accessed: 2025-02-20}.


\bibitem[O'hara et~al\mbox{.}(2011)]%
        {o2011blended}
\bibfield{author}{\bibinfo{person}{Kenton O'hara}, \bibinfo{person}{Jesper Kjeldskov}, {and} \bibinfo{person}{Jeni Paay}.} \bibinfo{year}{2011}\natexlab{}.
\newblock \showarticletitle{Blended interaction spaces for distributed team collaboration}.
\newblock \bibinfo{journal}{\emph{ACM Transactions on Computer-Human Interaction (TOCHI)}} \bibinfo{volume}{18}, \bibinfo{number}{1} (\bibinfo{year}{2011}), \bibinfo{pages}{1--28}.
\newblock


\bibitem[Paolo et~al\mbox{.}(2024)]%
        {paolo2024call}
\bibfield{author}{\bibinfo{person}{Giuseppe Paolo}, \bibinfo{person}{Jonas Gonzalez-Billandon}, {and} \bibinfo{person}{Bal{\'a}zs K{\'e}gl}.} \bibinfo{year}{2024}\natexlab{}.
\newblock \showarticletitle{A call for embodied AI}.
\newblock \bibinfo{journal}{\emph{arXiv preprint arXiv:2402.03824}} (\bibinfo{year}{2024}).
\newblock


\bibitem[Puig et~al\mbox{.}(2020)]%
        {puig2020watch}
\bibfield{author}{\bibinfo{person}{Xavier Puig}, \bibinfo{person}{Tianmin Shu}, \bibinfo{person}{Shuang Li}, \bibinfo{person}{Zilin Wang}, \bibinfo{person}{Yuan-Hong Liao}, \bibinfo{person}{Joshua~B Tenenbaum}, \bibinfo{person}{Sanja Fidler}, {and} \bibinfo{person}{Antonio Torralba}.} \bibinfo{year}{2020}\natexlab{}.
\newblock \showarticletitle{Watch-and-help: A challenge for social perception and human-ai collaboration}.
\newblock \bibinfo{journal}{\emph{arXiv preprint arXiv:2010.09890}} (\bibinfo{year}{2020}).
\newblock


\bibitem[{Rewatch.com}(2025)]%
        {Rewatch2025}
\bibfield{author}{\bibinfo{person}{{Rewatch.com}}.} \bibinfo{year}{2025}\natexlab{}.
\newblock \bibinfo{booktitle}{\emph{Rewatch}}.
\newblock
\urldef\tempurl%
\url{https://rewatch.com/}
\showURL{%
\tempurl}


\bibitem[Samrose et~al\mbox{.}(2021)]%
        {samrose2021meetingcoach}
\bibfield{author}{\bibinfo{person}{Samiha Samrose}, \bibinfo{person}{Daniel McDuff}, \bibinfo{person}{Robert Sim}, \bibinfo{person}{Jina Suh}, \bibinfo{person}{Kael Rowan}, \bibinfo{person}{Javier Hernandez}, \bibinfo{person}{Sean Rintel}, \bibinfo{person}{Kevin Moynihan}, {and} \bibinfo{person}{Mary Czerwinski}.} \bibinfo{year}{2021}\natexlab{}.
\newblock \showarticletitle{Meetingcoach: An intelligent dashboard for supporting effective \& inclusive meetings}. In \bibinfo{booktitle}{\emph{Proceedings of the 2021 CHI Conference on Human Factors in Computing Systems}}. \bibinfo{pages}{1--13}.
\newblock


\bibitem[Sayadi et~al\mbox{.}(2024)]%
        {sayadi2024feeling}
\bibfield{author}{\bibinfo{person}{Niloofar Sayadi}, \bibinfo{person}{Sadie Co}, {and} \bibinfo{person}{Diego Gomez-Zara}.} \bibinfo{year}{2024}\natexlab{}.
\newblock \showarticletitle{" Feeling that I was Collaborating with Them": A 20 years Systematic Literature Review of Social Virtual Reality Leveraging Collaboration}.
\newblock \bibinfo{journal}{\emph{arXiv preprint arXiv:2412.20266}} (\bibinfo{year}{2024}).
\newblock


\bibitem[Tang et~al\mbox{.}(2023)]%
        {tang2023perspectives}
\bibfield{author}{\bibinfo{person}{John~C Tang}, \bibinfo{person}{Kori Inkpen}, \bibinfo{person}{Sasa Junuzovic}, \bibinfo{person}{Keri Mallari}, \bibinfo{person}{Andrew~D Wilson}, \bibinfo{person}{Sean Rintel}, \bibinfo{person}{Shiraz Cupala}, \bibinfo{person}{Tony Carbary}, \bibinfo{person}{Abigail Sellen}, {and} \bibinfo{person}{William~AS Buxton}.} \bibinfo{year}{2023}\natexlab{}.
\newblock \showarticletitle{Perspectives: creating inclusive and equitable hybrid meeting experiences}.
\newblock \bibinfo{journal}{\emph{Proceedings of the ACM on Human-Computer Interaction}} \bibinfo{volume}{7}, \bibinfo{number}{CSCW2} (\bibinfo{year}{2023}), \bibinfo{pages}{1--25}.
\newblock


\bibitem[van~der Land et~al\mbox{.}(2011)]%
        {vanderLand2011ModelingMetaverse}
\bibfield{author}{\bibinfo{person}{Sarah~F. van~der Land}, \bibinfo{person}{Alexander~P. Schouten}, \bibinfo{person}{Bart~J. van~den Hooff}, {and} \bibinfo{person}{Frans Feldberg}.} \bibinfo{year}{2011}\natexlab{}.
\newblock \showarticletitle{Modeling the Metaverse: A Theoretical Model of Effective Team Collaboration in 3D Virtual Environments}.
\newblock \bibinfo{journal}{\emph{Journal of Virtual Worlds Research}} \bibinfo{volume}{4}, \bibinfo{number}{3} (\bibinfo{year}{2011}), \bibinfo{pages}{1--17}.
\newblock
\href{https://doi.org/10.4101/jvwr.v4i3.6126}{doi:\nolinkurl{10.4101/jvwr.v4i3.6126}}


\bibitem[Wan et~al\mbox{.}(2024)]%
        {Wan2024BuildingLLM}
\bibfield{author}{\bibinfo{person}{Hongyu Wan}, \bibinfo{person}{Jinda Zhang}, \bibinfo{person}{Abdulaziz~Arif Suria}, \bibinfo{person}{Bingsheng Yao}, \bibinfo{person}{Dakuo Wang}, \bibinfo{person}{Yolanda Coady}, {and} \bibinfo{person}{Maja Prpa}.} \bibinfo{year}{2024}\natexlab{}.
\newblock \showarticletitle{Building LLM-based AI Agents in Social Virtual Reality}. In \bibinfo{booktitle}{\emph{Proceedings of the 2024 CHI Conference on Human Factors in Computing Systems}}. \bibinfo{publisher}{Association for Computing Machinery}.
\newblock
\href{https://doi.org/10.1145/3613905.3651026}{doi:\nolinkurl{10.1145/3613905.3651026}}


\bibitem[Wang et~al\mbox{.}(2019)]%
        {wang2019exploring}
\bibfield{author}{\bibinfo{person}{Isaac Wang}, \bibinfo{person}{Jesse Smith}, {and} \bibinfo{person}{Jaime Ruiz}.} \bibinfo{year}{2019}\natexlab{}.
\newblock \showarticletitle{Exploring virtual agents for augmented reality}. In \bibinfo{booktitle}{\emph{Proceedings of the 2019 CHI Conference on Human Factors in Computing Systems}}. \bibinfo{pages}{1--12}.
\newblock


\bibitem[Yan et~al\mbox{.}(2023)]%
        {yan2023inherent}
\bibfield{author}{\bibinfo{person}{He Yan}, \bibinfo{person}{Xinyao Hu}, \bibinfo{person}{Xiangpeng Wan}, \bibinfo{person}{Chengyu Huang}, \bibinfo{person}{Kai Zou}, {and} \bibinfo{person}{Shiqi Xu}.} \bibinfo{year}{2023}\natexlab{}.
\newblock \showarticletitle{Inherent limitations of LLMs regarding spatial information}.
\newblock \bibinfo{journal}{\emph{arXiv preprint arXiv:2312.03042}} (\bibinfo{year}{2023}).
\newblock


\bibitem[Zhao and Baghaei(2024)]%
        {zhao2024exploring}
\bibfield{author}{\bibinfo{person}{Yinshu Zhao} {and} \bibinfo{person}{Nilufar Baghaei}.} \bibinfo{year}{2024}\natexlab{}.
\newblock \showarticletitle{Exploring AI Interaction Modalities in Virtual Environments and Its Impact}.
\newblock \bibinfo{journal}{\emph{The Impact of Artificial Intelligence on Societies}} (\bibinfo{year}{2024}), \bibinfo{pages}{41}.
\newblock


\end{thebibliography}

\end{document}